\documentclass{article}
\usepackage{spconf,amsmath,graphicx,hyperref}
\usepackage{amssymb}
\usepackage[colorinlistoftodos,prependcaption,backgroundcolor=black!5!white,bordercolor=red]{todonotes}

\title{JEPA-kMMVAE: CSI Prediction for Time-Varying Wireless Channels}
\name{%
    \begin{tabular}{@{}c@{}}
        Efstathios Chatziloizos$^{1,2}$,
        Konstantinos Vandikas$^{1}$,
        Aneta Vulgarakis Feljan$^{1}$ \\
        Zheng Chen$^{3}$ and Nikolaos Pappas$^{2}$
    \end{tabular}%
    \thanks{\fontsize{9}{11}\selectfont
    This work was supported by Ericsson Research and the
    Wallenberg AI, Autonomous Systems, and Software Program
    (WASP) funded by the Knut and Alice Wallenberg Foundation.
    The work of N. Pappas has been supported in part by ELLIIT
    and the European Union (6G-LEADER, 101192080).}
}

\address{%
    \fontsize{10}{12}\selectfont
    \begin{tabular}{@{}c@{}}
        $^{1}$ Ericsson Research, Stockholm, Sweden \\
        $^{2}$ Department of Computer and Information Science,
        Link\"oping University, Link\"oping, Sweden \\
        $^{3}$ Department of Electrical Engineering,
        Link\"oping University, Link\"oping, Sweden \\
        \texttt{\{efstathios.chatziloizos, konstantinos.vandikas, aneta.vulgarakis\}@ericsson.com} \\
        \texttt{\{efstathios.chatziloizos, zheng.chen, nikolaos.pappas\}@liu.se}
    \end{tabular}%
}
\begin{document}
%
\maketitle

\begin{abstract}
This paper considers channel state information (CSI)
prediction using neural networks for time-varying
multi-antenna channels from noisy historical observations.
The main challenge is to exploit temporal correlation
while distinguishing channel evolution from observation
noise. We propose JEPA-$k$MMVAE, which combines a
joint-embedding predictive architecture with a
$k$-memory Markov variational autoencoder to predict
future CSI through temporally structured latent
representations. The framework supports direct prediction
and residual corrections relative to the latest latent
state and noisy CSI observation.
Simulation results show that the residual configuration
outperforms the evaluated baselines at moderate and high
input CSI signal-to-noise ratios, while sequence baselines
remain more accurate under stronger noise.
\end{abstract}
\section{Introduction}
\label{sec:introduction}

Accurate channel state information (CSI) is essential for coherent detection and precoding in multi-antenna wireless systems. In time-varying propagation environments, however, CSI acquired from pilot observations may become outdated before it is used. Channel prediction addresses this mismatch by inferring future channel realizations from historical observations. This connects CSI prediction to the broader goals of
timeliness and task relevance in next-generation
wireless networks~\cite{popovski2022time,kountouris2021semantics, semanticsTCOMinvited}.
The central challenge is to exploit temporal dependence
while distinguishing channel evolution from noise
in the available observations.

For time-varying channel estimation, B\"ock \emph{et al.}
introduced the $k$-memory Markov variational autoencoder
($k$-MMVAE), combining Markov latent dynamics with
finite-memory inference and decoding~\cite{bock2023kmmvae}.
Our subsequent work adapted this structure to
context-aware CSI compression~\cite{chatziloizos2026context}.
For channel prediction, Jiang \emph{et al.} used
Transformers to forecast future channels in
parallel~\cite{jiang2022transformer}, while WiFo
introduced masked pretraining over heterogeneous
space--time--frequency CSI~\cite{liu2025wifo}.
Zhou \emph{et al.} combined predictive foundation-model
priors with current pilots for channel
estimation~\cite{zhou2026predictive}.
Here, we extend temporally structured variational
representations to one-step prediction from noisy
history, without an observation at the target time.

Our approach builds on two complementary ideas:
Joint-embedding predictive architectures (JEPAs)
predict target representations from context
representations~\cite{assran2023ijepa}, motivating
latent supervision alongside the CSI-domain objective.
Residual learning, exemplified by the denoising
convolutional neural network (DnCNN)~\cite{zhang2017dncnn},
suggests learning a correction to an available
observation. For prediction, the correction accounts for
observation noise and channel evolution.

In this paper, we introduce JEPA-$k$MMVAE to predict
the next noise-free CSI snapshot from noisy history.
The challenge is to exploit temporal dependence while
separating channel evolution from measurement noise.
The framework combines a pretrained $k$-MMVAE
representation, a temporal latent predictor, and a
future-CSI decoder with direct and residual variants,
alongside a clean target branch for latent supervision
during training.
Our contribution is the extension of temporally
structured variational channel representations to
causal future-CSI prediction.
We compare both variants against direct CSI baselines
and examine how observation quality affects their
relative performance.

\section{System Model and Prediction Problem}
\label{sec:system_model}

\begin{figure}[t]
    \centering
    \includegraphics[width=\columnwidth]{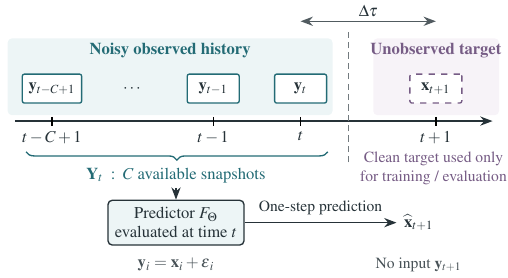}
    \caption{Causal CSI prediction from $C$ noisy observations. The predictor maps the available history $\mathbf Y_t$ to the next clean snapshot estimate $\widehat{\mathbf x}_{t+1}$. The future target is available only for training and evaluation.}
    \label{fig:system_prediction}
\end{figure}

Figure~\ref{fig:system_prediction} illustrates the causal prediction setting, where a finite history of noisy CSI measurements is used to predict the next clean channel snapshot one sampling interval ahead.

\subsection{Time-Varying Channel Observations}
\label{subsec:observations}

We consider a base station (BS) with $R$ antennas communicating with a set of single-antenna user equipments (UEs). For notational simplicity, we focus on one UE and omit its index in the rest of the paper. The channel between the BS and the UE is denoted by a vector $\mathbf h_t\in\mathbb C^R$, which corresponds to the channel coefficients sampled at time $t\Delta\tau$, where $\Delta\tau$ is the sampling interval. 
The narrowband formulation applies per subcarrier
in an orthogonal frequency-division multiplexing (OFDM) system,
without accounting for the inter-subcarrier correlation. Successive channel snapshots are correlated over time due to the evolution of the propagation environment and terminal position. 

In practical wireless systems, CSI is estimated from
pilot signals and is affected by estimation errors
and measurement noise. The CSI available for subsequent
processing therefore differs from the underlying
true channel.
For a window of $C$ consecutive snapshots indexed by
$i=t-C+1,\ldots,t$, we model the observed CSI as
\begin{equation}
    \mathbf r_i=\mathbf h_i+\mathbf n_i,
\end{equation}
where $\mathbf n_i$ is additive observation noise.
Conditioning on the clean channel history and selected
noise level, the noise vectors are independent with
$\mathbf n_i\sim\mathcal{CN}(\mathbf 0,\sigma_n^2\mathbf I_R)$,
using a common variance within each window~\cite{bock2023kmmvae}.
This effective Gaussian model represents uncertainty
in CSI acquisition.


For neural processing, we stack real and imaginary
components as
$\mathbf x_i=
[\operatorname{Re}\{\mathbf h_i\}^{\mathsf T},
 \operatorname{Im}\{\mathbf h_i\}^{\mathsf T}]^{\mathsf T}
\in\mathbb R^N$, where $N=2R$.
Applying the same stacking to $\mathbf r_i$ gives
\begin{equation}
    \mathbf y_i=\mathbf x_i+\boldsymbol\varepsilon_i,
    \qquad
    \boldsymbol\varepsilon_i\sim
    \mathcal N\!\left(
        \mathbf 0,\sigma_n^2/2\cdot\mathbf I_N
    \right).
    \label{eq:real_observation}
\end{equation}
This transformation preserves squared channel errors.

The  signal-to-noise ratio (SNR) of the input
CSI observations, defined using the latest snapshot
in the window, is $\gamma=\|\mathbf h_t\|_2^2/(R\sigma_n^2)$.
Earlier snapshots may have different instantaneous SNRs as channel power varies. 

\subsection{Causal CSI Prediction}
\label{subsec:prediction_problem}
Channel prediction exploits the temporal correlation
of wireless channels to infer future channel states
from past observations. Instead of relying solely on the most recent CSI estimate, a predictor can use a history
of CSI measurements to estimate the channel at a future
transmission time. Such predictive CSI can improve the timeliness of channel information and facilitate more effective beamforming, precoding, scheduling, and other channel-aware transmission decisions.

At time $t$, the predictor has access only to the
context $\mathbf Y_t=
[\mathbf y_{t-C+1},\ldots,\mathbf y_t]
\in\mathbb R^{N\times C}$.
The predictor crosses the observation boundary in Fig.~\ref{fig:system_prediction} through $\widehat{\mathbf x}_{t+1}=F_\Theta(\mathbf Y_t)$, where $\Theta$ denotes its parameters. 
The task therefore requires extrapolating channel evolution beyond the observed history while accounting for the observation noise. The context length $C$ is distinct from the local memory $k$ of the $k$-MMVAE introduced in Section~\ref{sec:proposed_framework}.

Under squared-error loss, the prediction risk is
\begin{equation}
    \mathcal R(F)=
    \mathbb E\!\left[
        \|\mathbf x_{t+1}-F(\mathbf Y_t)\|_2^2
    \right],
    \label{eq:prediction_risk}
\end{equation}
with expectation over channel evolution and observation
noise.
For finite second moments, its unrestricted minimizer
is the minimum mean-square error (MMSE) predictor,
$F^\star(\mathbf Y_t)
=\mathbb E[\mathbf x_{t+1}\mid\mathbf Y_t]$
\cite{kay1993estimation}.
JEPA-$k$MMVAE learns a parameterized predictor from
channel trajectories using temporal latent representations
and a future-CSI decoder.

For $M$ evaluated context--target pairs, the NMSE is
\begin{equation}
    \mathrm{NMSE}=
        \frac{1}{M}\sum_{m=1}^{M}
        \frac{\|\widehat{\mathbf x}^{(m)}
                  -\mathbf x^{(m)}\|_2^2}
             {\|\mathbf x^{(m)}\|_2^2},
    \label{eq:prediction_nmse}
\end{equation}
where $\mathbf x^{(m)}$ is the clean future target
of pair $m$. We report $10\log_{10}(\mathrm{NMSE})$
in dB.

\section{Proposed Prediction Framework}
\label{sec:proposed_framework}

\begin{figure*}[t]
    \centering
    \includegraphics[width=0.90\textwidth]
        {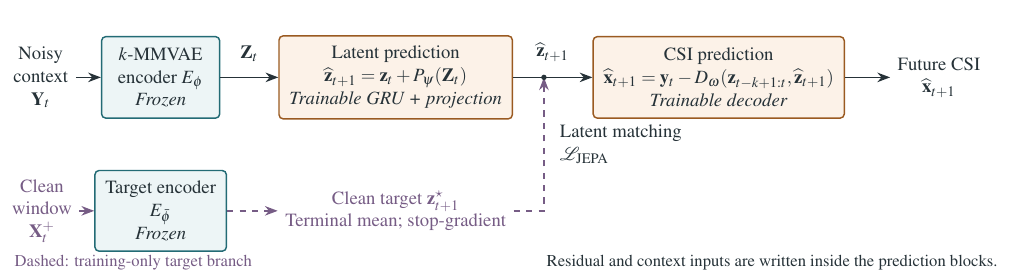}
    \caption{JEPA-$k$MMVAE residual prediction framework.
Noisy history drives latent prediction and residual
CSI decoding. The dashed branch supplies clean latent
targets during training; the CSI output is supervised
by the clean future snapshot.}
    \label{fig:proposed_approach}
\end{figure*}

JEPA-$k$MMVAE combines a pretrained $k$-MMVAE encoder
with a trainable latent predictor and future-CSI decoder.
Fig.~\ref{fig:proposed_approach} shows its residual
configuration with a denoising backbone.
Prediction uses only noisy history; the clean target
branch provides latent supervision during training.

\subsection{Temporally Structured Channel Representation}
\label{subsec:temporal_representation}

We adopt the $k$-MMVAE of~\cite{bock2023kmmvae}, also used for context-aware CSI compression in~\cite{chatziloizos2026context}. Each snapshot has a latent $\mathbf z_i\in\mathbb R^d$, with first-order prior $p_\eta(\mathbf z_i\mid\mathbf z_{i-1})$ and finite-memory inference model
\begin{align}
    q_\phi(\mathbf z_i\mid
            \mathbf z_{i-1},\mathbf y_{i-k:i})=\mathcal N\!\left(
        \boldsymbol\mu_{\phi,i},
        \operatorname{diag}(\boldsymbol\sigma_{\phi,i}^{2})
    \right).
    \label{eq:context_posterior}
\end{align}

The Gaussian parameters are learned functions of the conditioning variables; the prior is also learned Gaussian, with a standard normal at the first position. Position-specific network dependence is suppressed. Each window starts with a zero previous latent and zero-padded unavailable history. Thus, $k$ controls the explicit preceding observations, while the previous latent carries recursively obtained information.

The backbone is pretrained to reconstruct the clean terminal snapshot from noisy context using $k+1$ successive latents. Its objective combines squared reconstruction error with KL regularization against the Markov prior. After pretraining, it is frozen, and posterior means are propagated recursively to form
\begin{equation}
    \mathbf Z_t=E_\phi(\mathbf Y_t)
    =[\mathbf z_{t-C+1},\ldots,\mathbf z_t].
    \label{eq:context_embedding}
\end{equation}
Each $\mathbf z_i$ is the mean
in~\eqref{eq:context_posterior}, evaluated at the preceding deterministic latent. Propagating these means avoids latent sampling noise during prediction.

\subsection{Latent Prediction and Residual CSI Decoding}
\label{subsec:latent_prediction}

A gated recurrent unit (GRU)~\cite{cho2014gru} processes all $C$ context latents. A linear projection of its final hidden state produces the increment $P_\psi(\mathbf Z_t)$, giving
\begin{equation}
    \widehat{\mathbf z}_{t+1}
    =\mathbf z_t+P_\psi(\mathbf Z_t).
    \label{eq:latent_residual}
\end{equation}
This residual form learns a temporal correction to
the latest inferred latent. The future-CSI decoder
combines the prediction with the last $k$ context
latents to correct the latest noisy observation, giving
\begin{equation}
\begin{aligned}
    \widehat{\mathbf c}_{t+1}
        &=D_\omega(
            \mathbf z_{t-k+1:t},
            \widehat{\mathbf z}_{t+1}
        ),\\
    \widehat{\mathbf x}_{t+1}
        &=\mathbf y_t-\widehat{\mathbf c}_{t+1}.
\end{aligned}
\label{eq:csi_residual}
\end{equation}
Here, $D_\omega:\mathbb R^{(k+1)d}\rightarrow\mathbb R^N$
is a trainable head separate from the frozen
backbone decoder.

Inspired by residual denoising~\cite{zhang2017dncnn}, this formulation extends the correction to include channel evolution, yielding
\begin{equation}
    \mathbf c_{t+1}^{\star}
    =\mathbf y_t-\mathbf x_{t+1}
    =\boldsymbol\varepsilon_t
      -(\mathbf x_{t+1}-\mathbf x_t).
    \label{eq:residual_target}
\end{equation}
The decoder therefore addresses both observation noise and temporal change. Their relative importance depends on observation quality and the channel variation over the prediction interval; the correction is not restricted to estimating additive noise. A zero correction recovers persistence, $\widehat{\mathbf x}_{t+1}=\mathbf y_t$, and both residual references are available at inference.

\begin{table*}[!t]
    \centering
    \caption{Prediction NMSE in dB (lower is better).
    Values for learned models are mean $\pm$ sample
    standard deviation over three runs.
    Both JEPA-$k$MMVAE variants use $d=64$.
    Bold indicates the lowest mean in each column.}
    \label{tab:prediction_results}

    \fontsize{9}{11}\selectfont
    \setlength{\tabcolsep}{4pt}
    \renewcommand{\arraystretch}{1.18}

    \begin{tabular*}{\textwidth}
        {@{\extracolsep{\fill}}l*{5}{c}@{}}
        \hline
        \textbf{Method}
        & \textbf{5 dB}
        & \textbf{10 dB}
        & \textbf{15 dB}
        & \textbf{20 dB}
        & \textbf{25 dB} \\
        \hline

        Persistence
        & $-4.75$
        & $-9.24$
        & $-12.94$
        & $-15.35$
        & $-16.50$ \\

        Linear extrapolation
        & $+2.01$
        & $-2.95$
        & $-7.84$
        & $-12.49$
        & $-16.53$ \\

        Learned linear
        & $-8.98 \pm 0.01$
        & $-11.32 \pm 0.01$
        & $-12.43 \pm 0.03$
        & $-12.85 \pm 0.03$
        & $-12.99 \pm 0.04$ \\

        GRU
        & $-9.69 \pm 0.01$
        & $\boldsymbol{-12.48 \pm 0.06}$
        & $-14.28 \pm 0.13$
        & $-15.10 \pm 0.17$
        & $-15.41 \pm 0.19$ \\

        LSTM
        & $-9.34 \pm 0.03$
        & $-11.73 \pm 0.06$
        & $-13.14 \pm 0.14$
        & $-13.75 \pm 0.18$
        & $-13.97 \pm 0.20$ \\

        Transformer
        & $\boldsymbol{-10.02 \pm 0.06}$
        & $-12.44 \pm 0.06$
        & $-13.72 \pm 0.07$
        & $-14.23 \pm 0.07$
        & $-14.41 \pm 0.07$ \\
        \hline

        JEPA-$k$MMVAE (direct)
        & $-7.79 \pm 0.02$
        & $-10.66 \pm 0.07$
        & $-12.81 \pm 0.07$
        & $-13.95 \pm 0.13$
        & $-14.39 \pm 0.22$ \\

        JEPA-$k$MMVAE (residual)
        & $-6.61 \pm 0.03$
        & $-10.89 \pm 0.02$
        & $\boldsymbol{-14.47 \pm 0.07}$
        & $\boldsymbol{-17.05 \pm 0.19}$
        & $\boldsymbol{-18.47 \pm 0.28}$ \\
        \hline
    \end{tabular*}
\end{table*}

\subsection{Clean Target Embeddings and Training Objective}
\label{subsec:prediction_training}

Following JEPA's representation-prediction
principle~\cite{assran2023ijepa}, we use the shifted
clean window
$\mathbf X_t^+=[\mathbf x_{t-C+2},\ldots,\mathbf x_{t+1}]$
to construct the target embedding
\begin{equation}
    \mathbf z_{t+1}^{\star}
    =\operatorname{sg}\!\left(
        [E_{\bar\phi}(\mathbf X_t^+)]_C
      \right),
    \label{eq:clean_target_embedding}
\end{equation}
where $[\cdot]_C$ selects the terminal latent and
$\operatorname{sg}$ denotes stop-gradient, preventing
gradient flow to the target branch.
The target encoder $E_{\bar\phi}$ is a frozen copy
of the context encoder $E_\phi$. Both use the same
latent-state initialization and propagate posterior
means. Clean future CSI is used to construct training
targets and is never an input to the prediction path.

For a mini-batch of $B$ context--target pairs, define
\begin{equation}
\begin{aligned}
    \mathcal L_{\mathrm{CSI}}
        &=\frac{1}{BN}\sum_{b=1}^{B}
          \|\widehat{\mathbf x}_{t+1}^{(b)}
             -\mathbf x_{t+1}^{(b)}\|_2^2,\\
    \mathcal L_{\mathrm{JEPA}}
        &=\frac{1}{Bd}\sum_{b=1}^{B}
          \|\widehat{\mathbf z}_{t+1}^{(b)}
             -\mathbf z_{t+1}^{\star(b)}\|_2^2.
\end{aligned}
\label{eq:prediction_losses}
\end{equation}
We train $\psi$ and $\omega$ by minimizing
$\lambda_{\mathrm{CSI}}\mathcal L_{\mathrm{CSI}}
+\lambda_{\mathrm{JEPA}}\mathcal L_{\mathrm{JEPA}}$.
With the backbone frozen, the context KL term
contributes no gradient to $\psi$ or $\omega$. The CSI loss supervises prediction accuracy, while the JEPA loss aligns the predicted and clean target representations. 
At inference, only~\eqref{eq:context_embedding}--\eqref{eq:csi_residual} are evaluated.

\section{Simulation Results}
\label{sec:results}

\subsection{Experimental Setup}

We use QuaDRiGa 3GPP~38.901 UMa channels at $2.1$~GHz
($50\%$ LOS, $80\%$ indoor), with $R=32$ antennas and
$100$ snapshots spaced by $\Delta\tau=0.5$~ms.
Speeds are constant per trajectory, with centers
$\{1,10,30\}$~m/s selected with probabilities
$\{0.70,0.25,0.05\}$ and clipped Gaussian jitter
($10\%$ relative standard deviation).
Training, validation, and test sets contain $8000$,
$800$, and $800$ trajectories, respectively.
Learned predictors use $C=8$ observations;
JEPA-$k$MMVAE uses $k=3$ and $d=64$.
Evaluation uses all $92$ rolling windows per test
trajectory, giving $73\,600$ predictions per SNR
with common noisy inputs.
NMSE follows~\eqref{eq:prediction_nmse}.

We compare against persistence,
$\widehat{\mathbf x}_{t+1}=\mathbf y_t$,
and linear extrapolation,
$\widehat{\mathbf x}_{t+1}=2\mathbf y_t-\mathbf y_{t-1}$,
which require no training.
The learned linear baseline applies an affine mapping
to the vectorized context $\mathbf Y_t$.
GRU and long short-term memory
(LSTM)~\cite{hochreiter1997lstm} predictors use two
recurrent layers with $128$ hidden units per layer,
followed by a fully connected output layer that maps
the final hidden state to future CSI.
The Transformer baseline~\cite{vaswani2017attention}
uses two encoder layers, four attention heads,
$128$-dimensional embeddings, feedforward width $256$,
and learned positional embeddings.
A fully connected output layer maps the final token
representation to future CSI.
All learned baselines operate directly on the noisy
CSI context and are trained with squared-error loss
against the clean next snapshot.

All learned predictors use training SNRs sampled
uniformly from $[0,25]$~dB and validation at $10$~dB.
Baselines use Adam (learning rate $10^{-3}$, batch
size $256$) for up to $200$ epochs. JEPA-$k$MMVAE trains only the latent
predictor and future-CSI decoder, keeping the pretrained
backbone frozen. In the direct variant, the decoder maps the predicted latent and the last $k$ context latents directly to
future CSI.
The residual variant uses the denoising backbone and both
residual connections described in
Section~\ref{sec:proposed_framework}.

\subsection{Results and Discussion}

In Table~\ref{tab:prediction_results}, Transformer and GRU
achieve the lowest mean NMSE at $5$ and $10$~dB,
respectively, outperforming both JEPA-$k$MMVAE variants.
Residual JEPA-$k$MMVAE attains the lowest mean at
$15$--$25$~dB.
Its advantage is modest at $15$~dB, but increases to
$1.70$ and $1.94$~dB over the strongest baseline at
$20$ and $25$~dB, respectively.

\begin{figure}[t]
    \centering
    \includegraphics[width=0.75\columnwidth]
        {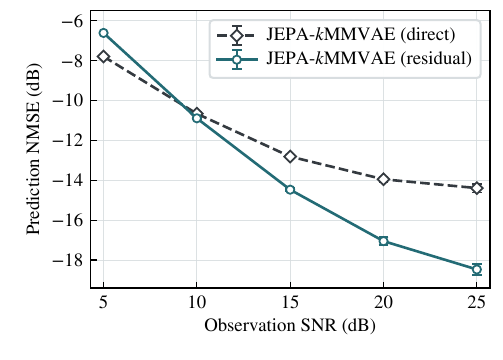}
    \caption{Direct and residual JEPA-$k$MMVAE prediction
    at $d=64$.
    Curves show mean CSI prediction NMSE; error bars denote
    $\pm$ one sample standard deviation over three runs.}
    \label{fig:prediction_configurations}
\end{figure}

Fig.~\ref{fig:prediction_configurations} compares the
two JEPA-$k$MMVAE variants.
Direct prediction is $1.18$~dB better at $5$~dB.
This ordering reverses by $10$~dB, and the residual
advantage grows to $4.08$~dB at $25$~dB. 

The correction target in~\eqref{eq:residual_target}
provides an interpretation of this crossover.
For slowly varying channels and weak observation noise,
the latest snapshot is a useful reference, allowing the
decoder to focus on the remaining channel change.
With stronger noise, the decoder must also remove the
corruption carried by this reference.
At $5$ and $10$~dB, the best sequence baseline achieves
lower mean NMSE than both variants.

\section{Conclusion}
\label{sec:conclusion}

We presented JEPA-$k$MMVAE for causal CSI prediction through
temporally structured latent representations.
The framework extends a pretrained Markov variational
encoder with latent forecasting and direct or residual
CSI decoding.
The evaluated residual configuration achieves the lowest
mean NMSE at moderate and high observation SNRs, while
learned baselines remain more accurate under stronger noise.
The direct--residual crossover highlights the importance
of observation reliability when using the latest snapshot
as a prediction reference.
Future work will investigate noise-aware combinations
of direct and residual prediction and extend the
framework to multi-step channel forecasting.

\bibliographystyle{IEEEbib}
\bibliography{refs}

\end{document}